\documentclass[traditabstract]{aa}

\usepackage{graphicx}
\usepackage{xspace}
\usepackage{natbib}
\usepackage{url}
\usepackage[xindy, toc, hyperfirst=false, nolist, nostyles, sanitize={name=false,description=false,symbol=false}]{glossaries}
\bibpunct{(}{)}{;}{a}{}{,}

\newcommand{\astropy}{\texttt{astropy}\xspace}

\begin{document}

% Reading the glossary
% Glossary for the Astropy papers
\newglossaryentry{astropy}{name=astropy, text=astropy}
\newglossaryentry{numpy}{name=NumPy, text=NumPy, description=Numerical Python, first=NumPy \citep{oliphant2006guide,van2011numpy}}
\newglossaryentry{scipy}{name=SciPy, text=SciPy, description={Scientific Python \cite{Jones:2001fk}}, first=SciPy \citep{jones2001scipy}}
\newglossaryentry{matplotlib}{name=Matplolib, text=Matplotlib}
\newglossaryentry{ipython}{name=IPython, text=IPython}
\newglossaryentry{python}{name=python, text=Python, description={Python programming language}}
\newglossaryentry{pandas}{name=pandas, text=pandas}
\newglossaryentry{wcs}{name=wcs, text=WCS, first=World Coordinate System WCS}

\titlerunning{Astropy}
\authorrunning{The Astropy Collaboration}

\title{Astropy: A Community Python Package for Astronomy}

\author{
The Astropy Collaboration
  \and
% Coordination committee
Thomas P. Robitaille\inst{\ref{inst:mpia}}  % confirmed
  \and
Erik J. Tollerud\inst{\ref{inst:yale}, \ref{inst:hubble}}  % confirmed
  \and
Perry Greenfield\inst{\ref{inst:stsci}}  % confirmed
  \and
% Developers (other than coordination committee).
Michael Droettboom\inst{\ref{inst:stsci}}  % confirmed
  \and
Erik Bray\inst{\ref{inst:stsci}}  % confirmed
  \and
Tom Aldcroft\inst{\ref{inst:cfa}}  % confirmed
  \and
Matt Davis\inst{\ref{inst:stsci}}  % confirmed
  \and
Adam Ginsburg\inst{\ref{inst:colorado}}  % confirmed
  \and
Adrian M. Price-Whelan\inst{\ref{inst:columbia}}  % confirmed
  \and
Wolfgang E. Kerzendorf\inst{\ref{inst:toronto}}  % confirmed
  \and
Alexander Conley\inst{\ref{inst:colorado}}  % confirmed
  \and
Neil Crighton\inst{\ref{inst:mpia}}  % confirmed
  \and
Kyle Barbary\inst{\ref{inst:argonne}}  % confirmed
  \and
Demitri Muna\inst{\ref{inst:osu}}  % confirmed
  \and
Henry Ferguson\inst{\ref{inst:stsci}}  % confirmed
  \and
Fr\'ed\'eric Grollier\inst{\ref{inst:freelance}}  % confirmed
  \and
Madhura M. Parikh\inst{\ref{inst:surat}}  % confirmed
  \and
Prasanth H. Nair\inst{\ref{inst:freelance}}  % confirmed
  \and
Hans M. G\"unther\inst{\ref{inst:cfa}}  % confirmed
  \and
Christoph Deil\inst{\ref{inst:mpik}}  % confirmed
  \and
Julien Woillez\inst{\ref{inst:eso_garching}}  % confirmed
  \and
Simon Conseil\inst{\ref{inst:oamp}}  % confirmed
  \and
Roban Kramer\inst{\ref{inst:eth}}  % confirmed
  \and
James E. H. Turner\inst{\ref{inst:gemini_s}}  % confirmed
  \and
Leo Singer\inst{\ref{inst:ligo}}  % confirmed
  \and
Ryan Fox\inst{\ref{inst:freelance}}  % confirmed
  \and
Benjamin A. Weaver\inst{\ref{inst:nyu}}  % confirmed
  \and
Victor Zabalza\inst{\ref{inst:mpik}}  % confirmed
  \and
Zachary I. Edwards\inst{\ref{inst:louisiana}}  % confirmed
  \and
% Other contributors, alphabetical, including coordination meeting
% participants
K. Azalee Bostroem\inst{\ref{inst:stsci}}  % confirmed
  \and
D. J. Burke\inst{\ref{inst:cfa}}  % confirmed
  \and
Andrew R. Casey\inst{\ref{inst:stromlo}}  % confirmed
  \and
Steven M. Crawford\inst{\ref{inst:saao}}  % confirmed
  \and
Nadia Dencheva\inst{\ref{inst:stsci}}  % confirmed
  \and
Justin Ely\inst{\ref{inst:stsci}}  % confirmed
  \and
Tim Jenness\inst{\ref{inst:jac},\ref{inst:cornell}}  % confirmed
  \and
Kathleen Labrie\inst{\ref{inst:gemini_n}}  % confirmed
  \and
Pey Lian Lim\inst{\ref{inst:stsci}}  % confirmed
  \and
Francesco Pierfederici\inst{\ref{inst:stsci}}  % confirmed
  \and
Andrew Pontzen\inst{\ref{inst:oxford},\ref{inst:ucl}}  % confirmed
  \and
Andy Ptak\inst{\ref{inst:gsfc}}  % confirmed
  \and
Brian Refsdal\inst{\ref{inst:cfa}}  % confirmed
  \and
Mathieu Servillat\inst{\ref{inst:saclay},\ref{inst:cfa}}  % confirmed
  \and
Ole Streicher\inst{\ref{inst:leibniz}}  % confirmed
}

\institute{
  Max-Planck-Institut f\"{u}r Astronomie, K\"onigstuhl 17, Heidelberg 69117, Germany
  \label{inst:mpia}
    \and
  Department of Astronomy, Yale University, P.O. Box 208101, New Haven, CT 06510, USA
  \label{inst:yale}
    \and
  Hubble Fellow
  \label{inst:hubble}
    \and
  Space Telescope Science Institute, 3700 San Martin Drive, Baltimore, MD 21218, USA
  \label{inst:stsci}
    \and
  Harvard-Smithsonian Center for Astrophysics, 60 Garden Street, Cambridge, MA, 02138, USA
  \label{inst:cfa}
    \and
  Center for Astrophysics and Space Astronomy, University of Colorado, Boulder, CO 80309, USA
  \label{inst:colorado}
    \and
  Department of Astronomy, Columbia University, Pupin Hall, 550W 120th St., New York, NY 10027, USA
  \label{inst:columbia}
    \and
  Department of Astronomy and Astrophysics, University of Toronto, 50 Saint George Street, Toronto, ON M5S3H4, Canada
  \label{inst:toronto}
    \and
  Argonne National Laboratory, High Energy Physics Division, 9700 South Cass Avenue, Argonne, IL 60439, USA
  \label{inst:argonne}
    \and
  Department of Astronomy, Ohio State University, Columbus, OH 43210, USA
  \label{inst:osu}
    \and
  S.V.National Institute of Technology, Surat., India
  \label{inst:surat}
    \and
  Independent developer
  \label{inst:freelance}
    \and
  Max-Planck-Institut f\"{u}r Kernphysik, P.O. Box 103980, 69029 Heidelberg, Germany
  \label{inst:mpik}
    \and
  European Southern Observatory, Karl-Schwarzschild-Str. 2, 85748, Garching bei M\"{u}nchen, Germany
  \label{inst:eso_garching}
    \and
  Laboratoire d'Astrophysique de Marseille, OAMP, Universit\'e Aix-Marseille et CNRS,
Marseille, France
  \label{inst:oamp}
    \and
  ETH Z\"{u}rich, Institute for Astronomy, Wolfgang-Pauli-Strasse 27, Building HIT, Floor J, CH-8093 Zurich, Switzerland
  \label{inst:eth}
    \and
  Gemini Observatory, Casilla 603, La Serena, Chile
  \label{inst:gemini_s}
    \and
  LIGO Laboratory, California Institute of Technology, 1200 E. California Blvd., Pasadena, CA, 91125, USA
  \label{inst:ligo}
    \and
  Center for Cosmology and Particle Physics, New York University, New York, NY 10003, USA
  \label{inst:nyu}
    \and
  Department of Physics and Astronomy, Louisiana State University, Nicholson Hall, Baton Rouge, LA 70803, USA
  \label{inst:louisiana}
    \and
  Research School of Astronomy and Astrophysics, Australian National University, Mount Stromlo Observatory, via Cotter Road, Weston Creek ACT 2611, Australia
  \label{inst:stromlo}
    \and
  SAAO, P.O. Box 9, Observatory 7935, Cape Town, South Africa
  \label{inst:saao}
    \and
  Joint Astronomy Centre, 660 N.\ A`oh\=ok\=u Place, Hilo, HI 96720, USA
  \label{inst:jac}
    \and
  Department of Astronomy, Cornell University, Ithaca, NY 14853, USA
  \label{inst:cornell}
    \and
  Gemini Observatory, 670 N.\ A`oh\=ok\=u Place, Hilo, Hawaii 96720, USA
  \label{inst:gemini_n}
    \and
  Oxford Astrophysics, Denys Wilkinson Building, Keble Road, Oxford OX1 3RH, UK
  \label{inst:oxford}
    \and
  Department of Physics and Astronomy, University College London, London WC1E 6BT, UK
  \label{inst:ucl}
    \and
  NASA Goddard Space Flight Center, X-ray Astrophysics Lab Code 662, Greenbelt, MD 20771, USA
  \label{inst:gsfc}
    \and
  Laboratoire AIM, CEA Saclay, Bat. 709, 91191 Gif-sur-Yvette, France
  \label{inst:saclay}
    \and
  Leibniz-Institut f\"{u}r Astrophysik Potsdam (AIP), An der Sternwarte 16, 14482 Potsdam, Germany
  \label{inst:leibniz}
}

\abstract{
We present the first public version (v0.2) of the open-source and community-developed
Python package, Astropy. This package provides core astronomy-related
functionality to the community, including support for domain-specific file
formats such as Flexible Image Transport System (FITS) files, Virtual
Observatory (VO) tables, and common ASCII table formats, unit and physical
quantity conversions, physical constants specific to astronomy, celestial
coordinate and time transformations, world coordinate system (WCS) support,
generalized containers for representing gridded as well as tabular data, and a
framework for cosmological transformations and conversions. Significant
functionality is under active development, such as a model fitting framework,
VO client and server tools, and aperture and point spread function (PSF)
photometry tools. The core development team is actively making additions and
enhancements to the current code base, and we encourage anyone interested to
participate in the development of future Astropy versions.
}

\keywords{Methods: data analysis -- Methods: miscellaneous -- Virtual Observatory Tools}

\maketitle

\section{Introduction}

The Python programming language\footnote{\url{http://www.python.org}} has become
one of the fastest-growing programming languages in the astronomy community in
the last decade (see e.g. \citealt{greenfield11} for a recent review). While there have been a number of efforts to develop Python
packages for astronomy-specific functionality, these efforts have been
fragmented, and several dozens of packages have been developed across the
community with little or no coordination. This has led to duplication and a
lack of homogeneity across packages, making it difficult for users to install
all the required packages needed in an astronomer's toolkit. Because a number
of these packages depend on individual or small groups of developers, packages
are sometimes no longer maintained, or simply become unavailable, which is
detrimental to long-term research and reproducibility.

Motivated by these issues, the Astropy project was started in
2011 out of a desire to bring
together developers across the field of astronomy in order to coordinate the
development of a common set of Python tools for astronomers and
simplify the landscape of available packages. The project has grown rapidly,
and to date, over 200 individuals are signed up to the development mailing list for
the Astropy project.\footnote{
\url{https://groups.google.com/forum/?fromgroups#!forum/astropy-dev}}

One of the primary aims of the Astropy \textit{project} is to develop a core
\astropy \textit{package} that covers much of the astronomy-specific
functionality needed by researchers, complementing more general scientific
packages such as \gls{numpy} and \gls{scipy}, which are invaluable for
numerical array-based calculations and more general scientific algorithms (e.g.
interpolation, integration, clustering). In addition, the Astropy project
includes work on more specialized Python packages (which we call
\textit{affiliated packages}) that are not included in the core package for
various reasons: for some the functionality is in early stages of development
and is not robust; the license is not compatible with Astropy; the package
includes large files; or the functionality is mature, but too domain-specific
to be included in a core package.

The driving interface design philosophy behind the core package is that code
using \astropy should result in concise and easily readable code, even by those new to
Python. Typical operations should appear in code similar to how they would
appear if expressed in spoken or written language. Such an interface results in
code that is less likely to contain errors and is easily understood, enabling
astronomers to focus more of their effort on their science objectives rather
than interpreting obscure function or variable names or otherwise spending time
trying to understand the interface.

In this paper, we present the first public release (v0.2) of the \astropy
package. We provide an overview of the current capabilities
(\S\ref{sec:capabilities}), our development workflow (\S\ref{sec:workflow}),
and planned functionality (\S\ref{sec:future}). This paper is not intended to
provide a detailed documentation for the package (which is available
online\footnote{\url{http://docs.astropy.org}}), but is rather intended to give
an overview of the functionality and design.

\section{Capabilities}

This section provides a broad overview of the capabilities of the different
\astropy sub-packages, which covers units and unit conversions
(\S\ref{sec:units_main}), absolute dates and times (\S\ref{sec:time}),
celestial coordinates (\S\ref{sec:coordinates}), tabular and gridded data
(\S\ref{sec:table}), common astronomical file formats (\S\ref{sec:io}), world
coordinate system transformations (\S\ref{sec:wcs}), and cosmological utilities
(\S\ref{sec:cosmology}). We have illustrated each section with simple and
concise code examples, but for more details and examples, we refer the reader
to the online documentation.\footnotemark[3]

\label{sec:capabilities}

\subsection{Units, Quantities, and Physical Constants}

\label{sec:units_main}

The \texttt{astropy.units} sub-package provides support for physical units. It
originates from code in the \texttt{pynbody} package \citep{pynbody}, but has been
significantly enhanced in behavior and implementation (with the intent that the \texttt{pynbody} will eventually become interoperable with \texttt{astropy.units}). This sub-package can be
used to attach units to scalars and arrays, convert from one set of units to
another, define custom units, define equivalencies for units that are not
strictly the same (such as wavelength and frequency), and decompose units into
base units. Unit definitions are included in both the International System of
Units (SI) and the Centimeter-Gram-Second (CGS) systems, as well as a number of
astronomy- and astrophysics-specific units.

\subsubsection{Units}

\begin{figure}
\center
\caption{Quantity conversion using the \texttt{astropy.units} sub-package.\label{code:quantities}}
\vspace{0.1in}
\fbox{\includegraphics[scale=0.82,clip]{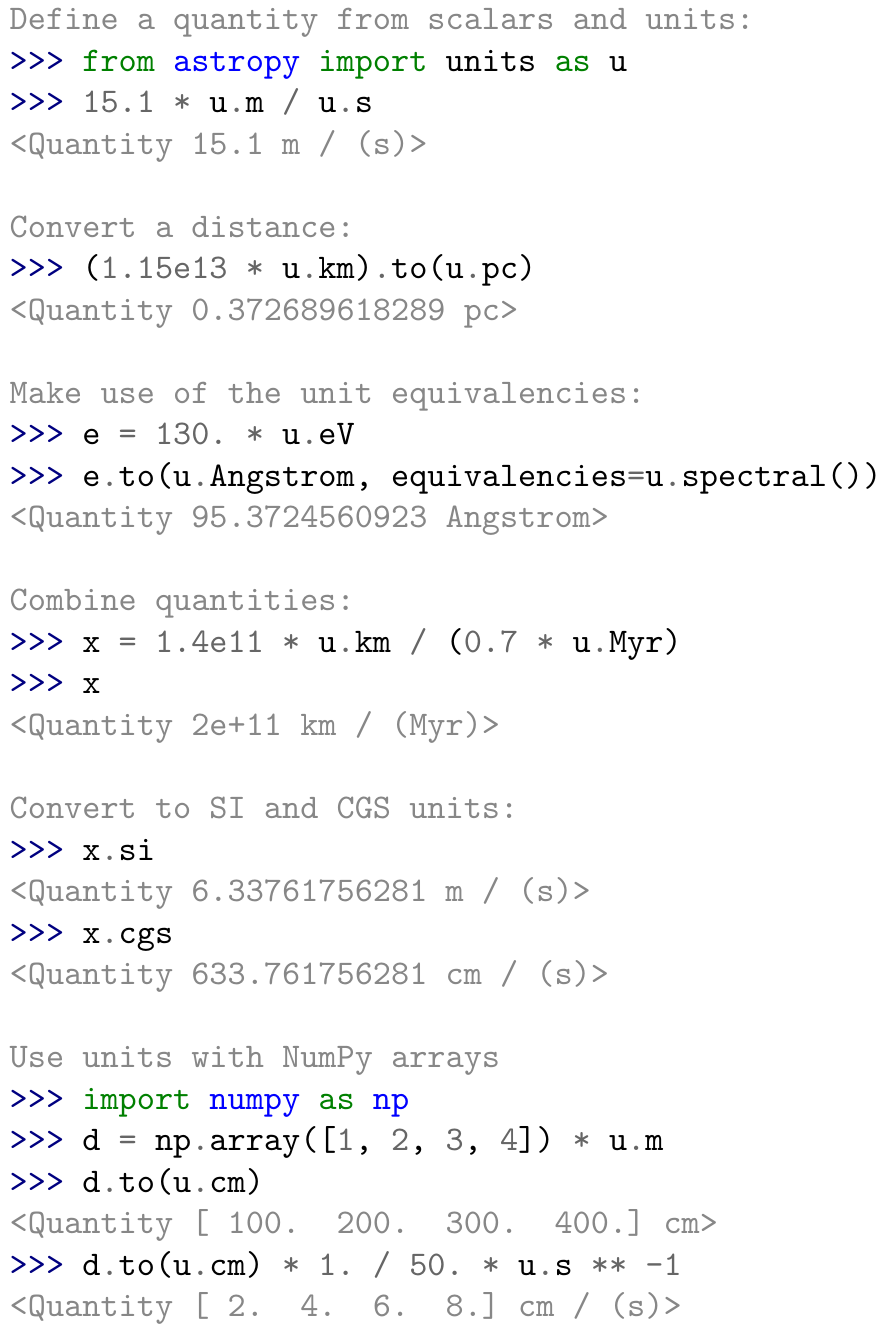}}
\end{figure}
\label{sec:units}

The \texttt{astropy.units} sub-package defines a \texttt{Unit} class to
represent base units, which can be manipulated without attaching them to
values, for example to determine the conversion factor from one set of units to
another. Users can also define their own units, either as standalone base units
or by composing other units together. It is also possible to decompose units
into their base units, or alternatively search for higher-level units that are
identical.

This sub-package includes the concept of ``equivalencies'' in units, which is
intended to be used where there exists an equation that provides a relationship
between two different physical quantities. A standard astronomical example is the
relationships between the frequency, wavelength and energy of a photon - it is
common practice to treat such units as equivalent even though they are not
strictly comparable. Such a conversion can be carried out in \texttt{astropy.units} by supplying an equivalency list
(see Figure~\ref{code:quantities}). The inclusion of these equivalencies is an important
improvement over existing unit-handling software, which typically does not
have this functionality. Equivalencies are also included for
monochromatic flux densities, which allows users to convert between
$F_\nu$ and $F_\lambda$, and users can easily implement their own
equivalencies.

There are multiple string representations for units used in the astronomy
community. The FITS Standard \citep{fits3} defines a unit standard, as well
as both the Centre de Donn\'ees astronomiques de Strasbourg (CDS)
\citep{ochsenbein2000cds} and NASA/Goddard's Office of Guest Investigator
Programs (OGIP) \citep{george1995ogip}. In addition, the International Virtual
Observatory Alliance (IVOA) has a forthcoming VOUnit standard
\citep{derriere2012vounit} in an attempt to resolve some of these differences.
Rather than choose one of these, \texttt{astropy.units} supports most of these
standards (OGIP support is planned for the next major release of \astropy),
and allows the user to select the appropriate one when reading and writing unit
string definitions to and from external file formats.

\subsubsection{Quantities and Physical Constants}

\label{sec:quantities}

While the previous section described the use of the \texttt{astropy.units} sub-package to manipulate
the units themselves, a more common use-case is to attach the units to
quantities, and use them together in expressions. The \texttt{astropy.units}
package allows units to be attached to Python scalars, or \gls{numpy} arrays,
producing \texttt{Quantity} objects. These objects support arithmetic with
other numbers and \texttt{Quantity} objects while preserving their units. For
multiplication and division, the resulting object will retain all units used in
the expression. The final object can then be converted to a specified set of
units or decomposed, effectively canceling and combining any equivalent units
and returning a \texttt{Quantity} object in some set of base units. This is
demonstrated in Figure~\ref{code:quantities}.

Using the \texttt{.to()} method, \texttt{Quantity} objects can easily be
converted to different units. The units must either be dimensionally
equivalent, or users should pass equivalencies through the
\texttt{equivalencies} argument (c.f. \S\ref{sec:units} or Figure~\ref{code:quantities}).
Since \texttt{Quantity} objects can operate with \gls{numpy} arrays, it is very simple
and efficient to convert the units on large datasets.

\begin{figure}
\center
\caption{Using the \texttt{astropy.constants} sub-package.\label{code:constants}}
\vspace{0.1in}
\fbox{\includegraphics[scale=0.82,clip]{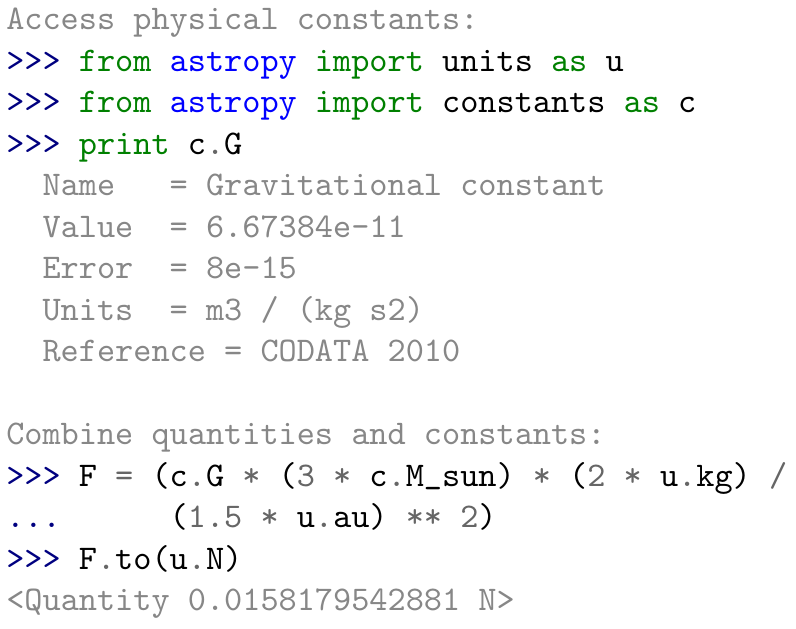}}
\end{figure}

The \texttt{Quantity} objects are used to define a number of useful
astronomical constants included in \texttt{astropy.constants}, each with an associated unit
(where applicable) and additional metadata describing their provenance and
uncertainties. These can be used along with \texttt{Quantity} objects to
provide a convenient framework for computing any quantity in astronomy.
Figure~\ref{code:constants} includes a simple example that shows how the
gravitational force between two bodies can be calculated in Newtons using
physical constants and user-specified quantities.

\subsection{Time}

\label{sec:time}

\begin{figure}
\center
\caption{Time representation and conversion using the \texttt{astropy.time}
sub-package.\label{code:time}}
\vspace{0.1in}
\fbox{\includegraphics[scale=0.82,clip]{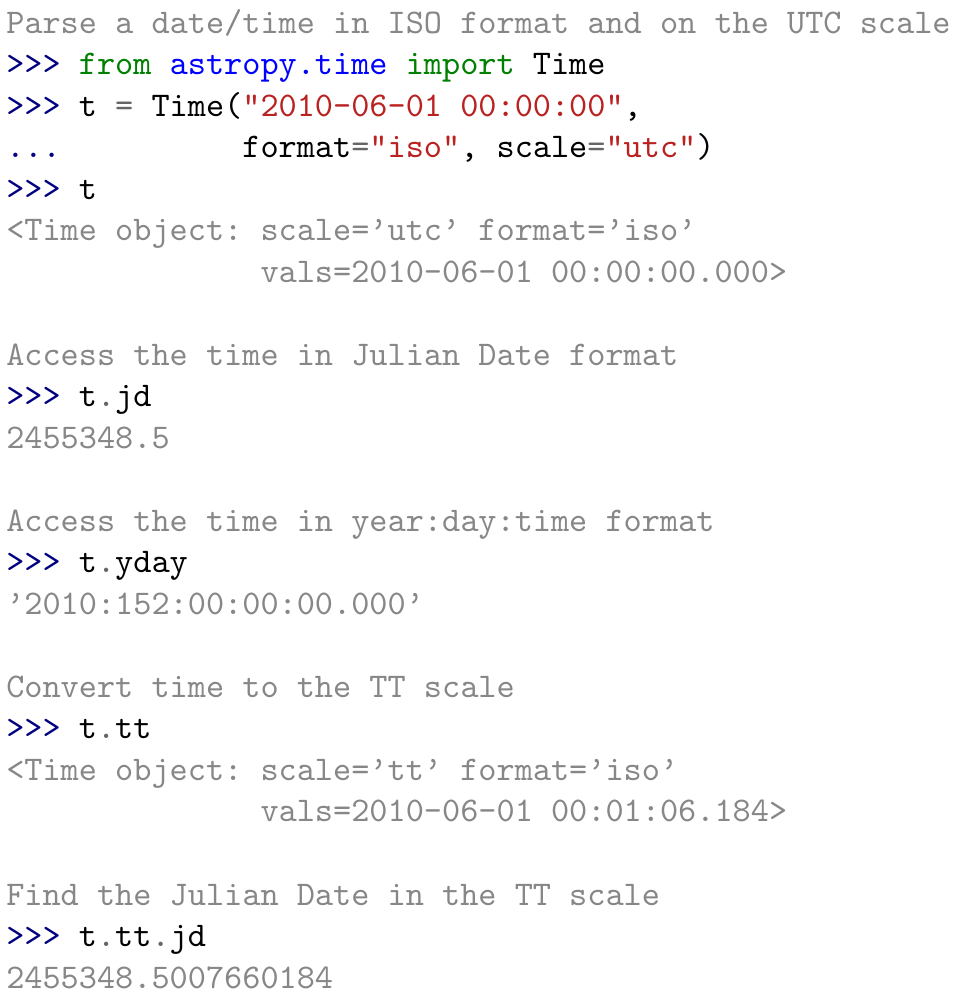}}
\end{figure}

The \texttt{astropy.time} package provides functionality for manipulating times
and dates. Specific emphasis is placed on supporting time scales (e.g. UTC,
TAI, UT1) and time formats or representations (e.g. JD, MJD, ISO 8601) that are
used in astronomy \citep{guinot88,kovalevsky01,sofa_wallace}. Examples of using this sub-package are provided in Figure
\ref{code:time}.

The most common way to use \texttt{astropy.time} is to create a \texttt{Time}
object by supplying one or more input time values as well as the time
format or representation and time scale of those values. The input time(s) can
either be a single scalar such as \verb|"2010-01-01 00:00:00"| or \verb|2455348.5|
or a sequence of such values; the format or representation specifies how to
interpret the input values, such as ISO, JD, or Unix time; and the scale
specifies the time standard used for the values, such as Coordinated Universal
Time (UTC), Terrestial Time (TT), or International Atomic Time (TAI). The full
list of available time scales is given in Table~\ref{tab:time_systems}. Many
of these formats and scales are used within astronomy, and it is especially
important to treat the different time scales properly when converting between
celestial coordinate systems. To facilitate this, the \texttt{Time} class makes
the conversion to a different format such as Julian Date straightforward, as
well as the conversion to a different time scale, for instance from UTC to TT.
We note that the \texttt{Time} class includes support for leap seconds
in the UTC time scale.

This package is based on a derived version of the Standards of Fundamental
Astronomy (SOFA) time and calendar
library\footnote{\url{http://www.iausofa.org}} \citep{sofa_wallace}.
Leveraging the robust and well-tested SOFA routines ensures that the
fundamental time scale conversions are being computed correctly. An important
feature of the SOFA time library which is supported by \texttt{astropy.time} is
that each time is represented as a pair of double-precision (64-bit)
floating-point values, which enables extremely high precision time
computations. Using two 64-bit floating-point values allows users to represent
times with a dynamic range of 30 orders of magnitude, providing for example times
accurate to better than a nanosecond over timescales of tens of Gyr. All time
scale conversions are done by vectorized versions of the SOFA routines using
Cython \citep{cython}, a Python package
that makes it easy to use C code in Python.

\begin{table}
\caption{Supported time scales for \texttt{astropy.time}\label{tab:time_systems}}
\center
\begin{tabular}{ll}
\hline
Scale  & Description \\
\hline
TAI    & International Atomic Time \\
TCB    & Barycentric Coordinate Time \\
TCG    & Geocentric Coordinate Time \\
TDB    & Barycentric Dynamical Time \\
TT     & Terrestrial Time \\
UT1    & Universal Time \\
UTC    & Coordinated Universal Time \\
\hline
\end{tabular}
\end{table}

\subsection{Celestial Coordinates}

\label{sec:coordinates}

An essential element of any astronomy workflow is the manipulation, parsing,
and conversion of astronomical coordinates. This functionality is provided in
Astropy by the \texttt{astropy.coordinates} sub-package. The aim of this
package is to provide a common application programming interface (API) for
Python astronomy packages that use coordinates, and to relieve users from
having to (re)implement extremely common utilities. To achieve this, it
combines API and implementation ideas from existing Python coordinates packages.
Some aspects, such as coordinate transformation approaches from
\texttt{kapteyn} \citep{kapteyn} and class structures resembling
\texttt{astropysics} \citep{astropysics}, have already been implemented.
Others, such as the frames of \texttt{palpy} \citep{palpy}
and \texttt{pyast} \citep{pyast} or the ephemeris system of
\texttt{pyephem} \citep{pyephem}, are still under design for \texttt{astropy}.
By combining the best aspects of these other packages, as well as
testing against them, \texttt{astropy.coordinates} seeks to provide a high-quality,
flexible Python coordinates library.

The sub-package has been designed to present a natural Python interface for
representing coordinates in computations, simplify input and output formatting,
and allow straightforward transformation between coordinate systems. It also
supports implementation of new or custom coordinate systems that work
consistently with the built-in systems. A future design goal is also to
seamlessly support arbitrarily large data sets.

\begin{figure}
\center
\caption{Celestial coordinate representation and conversion.\label{code:coords}}
\vspace{0.1in}
\fbox{\includegraphics[scale=0.82,clip]{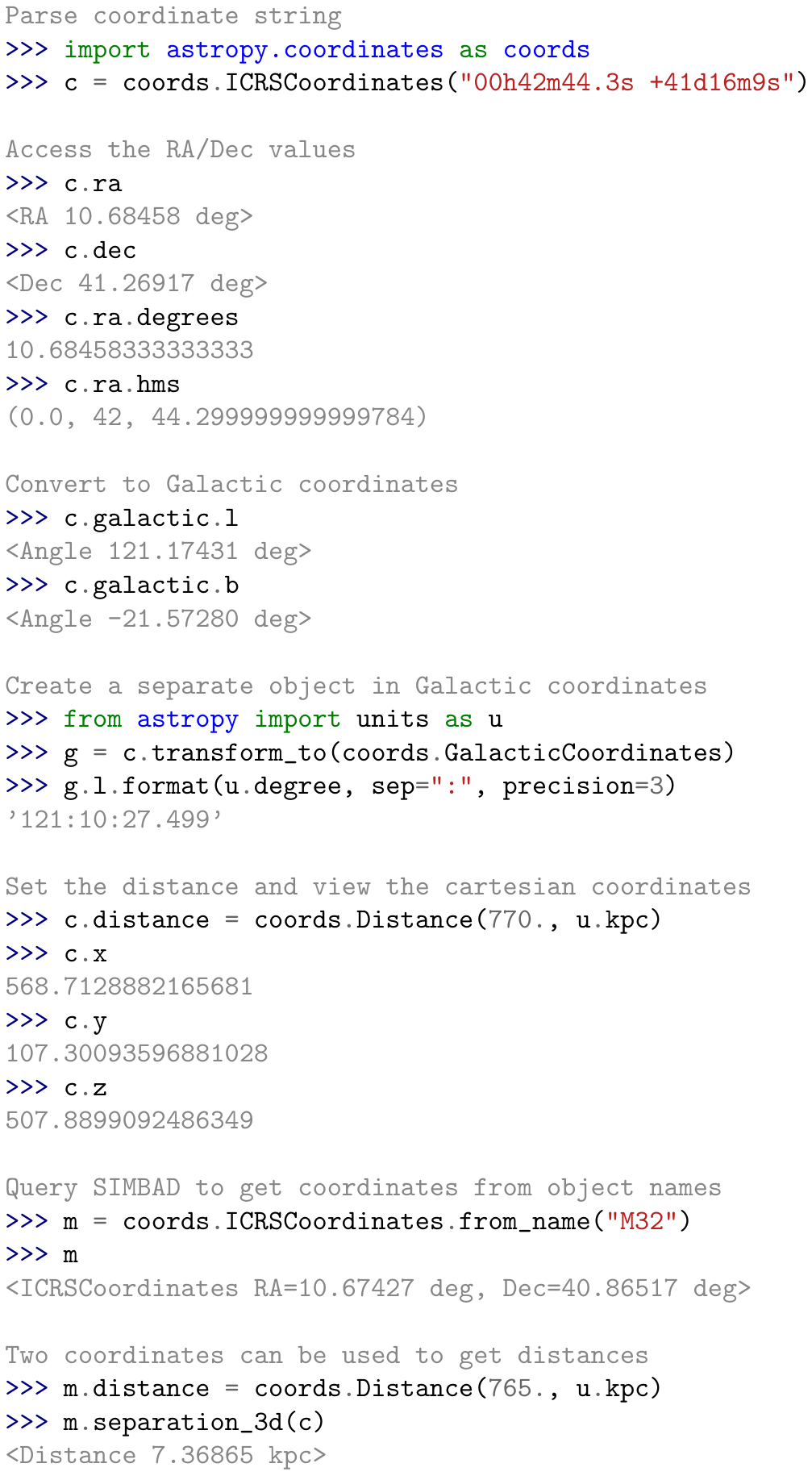}}
\label{fig:code_coordinates}
\end{figure}

To that end, Figure \ref{fig:code_coordinates} shows some typical usage
examples for \texttt{astropy.coordinates}. Coordinate objects are created using
standard Python object instantiation via a Python class named after the
coordinate system (e.g., \texttt{ICRSCoordinates}). Astronomical coordinates
may be expressed in a myriad of ways: the classes support string, numeric, and
tuple value specification through a sophisticated input parser. A design goal
of the input parser is to be able to determine the angle value and unit from
the input alone if a person can unambiguously determine them. For example, an
astronomer seeing the input string ``12h53m11.5123s'' would understand the
units to be in hours, minutes, and seconds, so this value is alone sufficient
to pass to the angle initializer. This functionality is built around the
\texttt{Angle} object, which can be instantiated and used on its own. It
provides additional functionality such as string formatting and mechanisms to
specify the valid bounds of an angle. As a convenience, it is also possible to
query the online SIMBAD\footnote{\url{http://simbad.u-strasbg.fr}} database to
resolve the name of a source (see Figure~\ref{fig:code_coordinates} for an
example showing how to find the ICRS coordinates of M32).

The coordinate classes represent different coordinate systems, and provide most
of the user-facing functionality for \texttt{astropy.coordinates}. The systems
provide customized initializers and appropriate formatting and representation
defaults. For some classes, they also contain added functionality specific to a
subset of systems, such as code to precess a coordinate to a new equinox. The
implemented systems include a variety of equatorial coordinate systems (ICRS,
FK4, and FK5), Galactic coordinates, and horizontal (Alt/Az) coordinates, and
modern (IAU 2006/200A) precession/nutation models for the relevant systems.
Coordinate objects can easily be transformed from one coordinate system to another: Figure \ref{fig:code_coordinates}
illustrates the most basic use of this functionality to convert a position on the sky from ICRS to
Galactic coordinates. Transformations are provided between all coordinate
systems built into version v0.2 of Astropy, with the exception of conversions
from celestial to horizontal coordinates. Future versions of Astropy will
include additional common systems, including
ecliptic systems, supergalactic coordinates, and all necessary intermediate
coordinate systems for the IAU 2000/2006 equatorial-to-horizontal mapping
\citep[e.g.,][]{soffel03, usnocircular179}.

A final significant feature of \texttt{astropy.coordinates} is support for
line-of-sight distances. While the term ``celestial coordinates'' can be taken
to refer to only on-sky angles, in \texttt{astropy.coordinates} a coordinate
object is conceptually treated as a point in three dimensional space. Users
have the option of specifying a line of sight distance to the object from the
origin of the coordinate system (typically the origin is the Earth or solar
system barycenter). These distances can be given in physical units or as
redshifts. The \texttt{astropy.coordinates} sub-package will in the latter case
transparently make use of the cosmological calculations in
\texttt{astropy.cosmology} (c.f. \S\ref{sec:cosmology}) for conversion to
physical distances. Figure \ref{fig:code_coordinates} illustrates an
application of this information in the form of computing three-dimensional
distances between two objects.

The \texttt{astropy.coordinates} sub-package was designed such that it should be easy for a user to add new
coordinate systems.
This flexibility is achieved in \texttt{astropy.coordinates} through the internal use of
a transformation graph, which keeps track of a network of
coordinate systems and the transformations between them. When a coordinate
object is to be transformed from one system into another, the package
determines the shortest path on the transformation graph to the new system and
applies the necessary sequence of transformations. Thus, implementing a new
coordinate system simply requires implementing one pair of transformations to and
from a system that is already connected to the transformation graph. Once this
pair is specified, \texttt{astropy.coordinates} can transform from that
coordinate system to any other in the graph.
An example of a user-defined system is provided in the
documentation,\footnote{\url{http://docs.astropy.org/en/v0.2.3/coordinates/sgr-example.html}}
illustrating the definition of a coordinate system useful for a specific
scientific task (Price-Whelan \& Johnston 2013, in prep).

\subsection{Tables and Gridded data}

\label{sec:table}

\begin{figure}
\center
\caption{Table input/output and manipulation using the \texttt{astropy.table}
sub-package.\label{code:tables}}
\vspace{0.1in}
\fbox{\includegraphics[scale=0.82,clip]{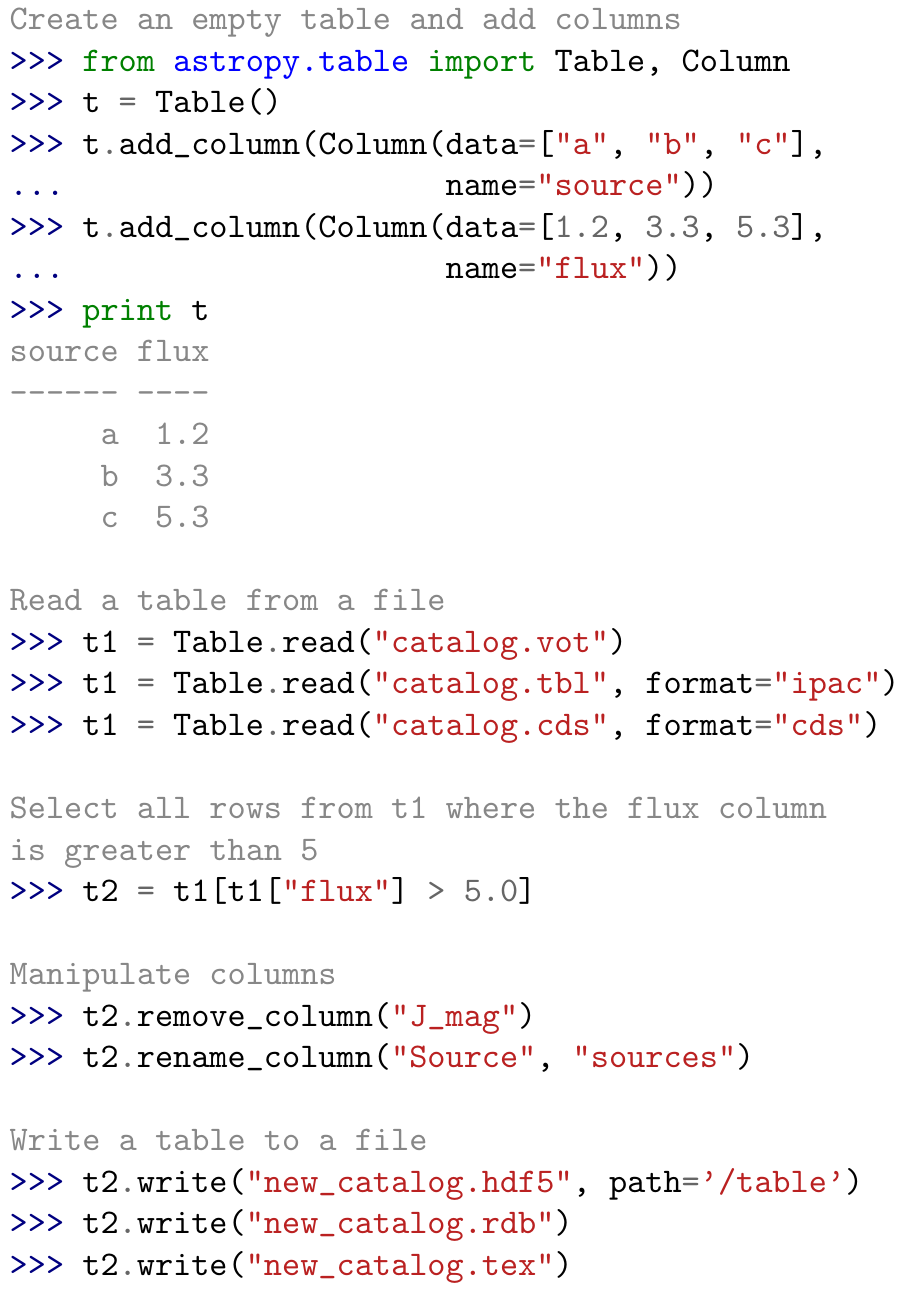}}
\end{figure}

Tables and $n$-dimensional data arrays are the most common forms of data
encountered in astronomy. The \gls{python} community has various solutions for
tables, such as \gls{numpy} structured arrays or \texttt{DataFrame} objects in
Pandas \citep{pandas} to name only a couple. For
n-dimensional data the \gls{numpy} \texttt{ndarray} is the most popular.

However, for use in astronomy all of these implementations lack some key
features. The data that is stored in arrays and tables often contains vital
metadata: the data is associated with units, and might also contain additional
arrays that either mask or provide additional attributes to each cell.
Furthermore, the data often includes a set of keyword-value pairs and comments (such as FITS headers).
Finally, the data comes in a plethora of astronomy specific formats (FITS,
specially formatted ASCII tables, etc.), which are not recognized by the
pre-existing packages.

The \texttt{astropy.table} and \texttt{astropy.nddata} sub-packages contain
classes (\texttt{Table} and \texttt{NDData}) that try to alleviate these
problems. They allow users to represent astronomical data in the form of tables
or n-dimensional gridded datasets, including all metadata. Examples of usage
of \texttt{astropy.table} are shown in Figure~\ref{code:tables}.

The \texttt{Table} class provides a high-level wrapper to \gls{numpy} structured
arrays, which are essentially arrays that have fields (or columns) with
heterogeneous data types, and any number of rows. \gls{numpy} structured arrays are
however difficult to modify, so the \texttt{Table} class makes it
easy for users to create a table from columns, add and remove columns or rows,
and mask values from the table. Furthermore, tables can be easily read from and
written to common file formats using the \texttt{Table.read} and
\texttt{Table.write} methods. These methods are connected to sub-packages in
\texttt{astropy.io} such as \texttt{astropy.io.ascii} (\S\ref{sec:ascii}) and
\texttt{astropy.io.votable} (\S\ref{sec:votable}), which allow ASCII and VO
tables to be seamlessly read or written respectively.

In addition to providing easy manipulation and input or output of table objects,
the \texttt{Table} class allows units to be specified for each column using the
\texttt{astropy.units} framework (\S\ref{sec:units_main}), and also allows the \texttt{Table} object to
contain arbitrary metadata (stored in \texttt{Table.meta}).

Similarly, the \texttt{NDData} class provides a way to store $n$-dimensional array
data easily and builds upon the \gls{numpy} \texttt{ndarray} class. The actual data is
stored in an \texttt{ndarray}, which allows for easy compatibility with
other scientific packages. In addition to keyword-value metadata, the
\texttt{NDData} class can store a boolean mask with the same dimensions as the
data, several sets of flags ($n$-dimensional arrays that store attributes for
each cell of the data array), uncertainties, units, and a transformation
between array-index coordinate system and other coordinate systems (c.f.
\S\ref{sec:wcs}). In addition, the \texttt{NDData} class intends to provide
methods to arithmetically combine the data in a meaningful way. \texttt{NDData}
is not meant for direct user interaction but more for providing a framework for
higher-level subclasses that can represent for example spectra or astronomical images.

\subsection{File Formats}

\label{sec:io}

\subsubsection{FITS}

\label{sec:fits}

\begin{figure}
\center
\caption{Accessing data in FITS format.\label{code:fits}}
\vspace{0.1in}
\fbox{\includegraphics[scale=0.82,clip]{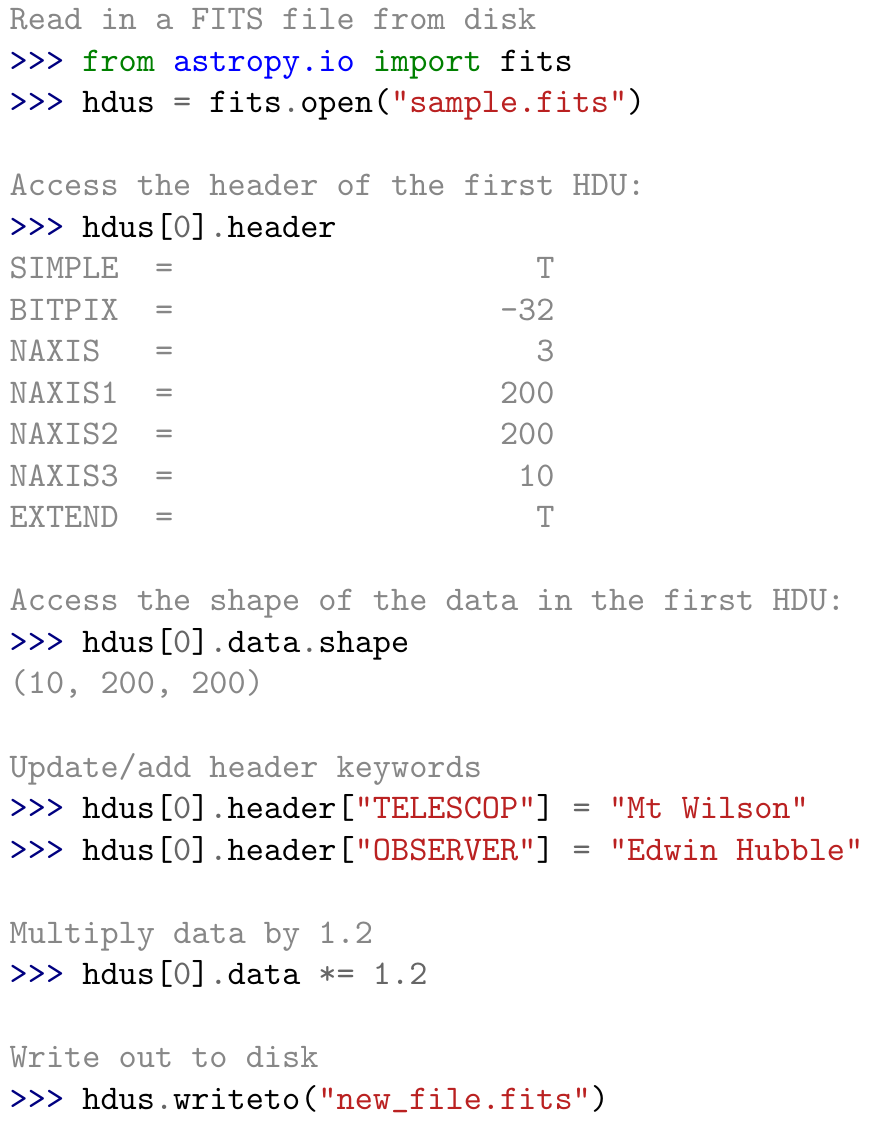}}
\end{figure}

Support for reading and writing FITS files is provided by the
\texttt{astropy.io.fits} sub-package, which at the time of writing is a direct
port of the
PyFITS\footnote{\url{http://www.stsci.edu/institute/software_hardware/}}
project \citep{barrett1999pyfits}. Users already familiar with PyFITS will
therefore feel at home with this package.

The \texttt{astropy.io.fits} sub-package implements all features from the FITS
standard \citep{fits3} such as images, binary tables, and ASCII tables, and
includes common compression algorithms. Header-data units (HDUs) are
represented by Python classes, with the data itself stored using NumPy arrays,
and with the headers stored using a \texttt{Header} class. Files can easily be
read and written, and once in memory can be easily modified. This
includes support for transparently reading from and writing to gzip-compressed
FITS files.
Figure~\ref{code:fits} shows a simple example of how to open an existing FITS
file, access and modify the header and data, and write a new file back to disk.

Creating new FITS files is also made simple. Since the code in this sub-package
has been developed over more than a decade, it has been made to work with an
extensive variety of FITS files, including ones that deviate from the FITS
standard. This includes support for deprecated formats such as
\texttt{GROUPS} HDUs as well as more obscure non-standard HDU types such as
\texttt{FITS} HDUs which allow encapsulating multiple FITS files within FITS
files. Support is also included for common but non-standard header conventions
such as \texttt{CONTINUE} cards and ESO \texttt{HIERARCH} cards.  Two
command-line utilities for working with FITS files are packaged with Astropy:
\texttt{fitscheck} can be used to validate FITS files against the standard.
\texttt{fitsdiff} can be used to compare two FITS files on a number of
criteria, and also includes a powerful API for programmatically comparing FITS
files.

Because the interface is exactly the same as that of PyFITS, users may directly
replace PyFITS with Astropy in existing code by changing import statements such as
\texttt{import pyfits} to \texttt{from astropy.io import fits as pyfits}
without any additional code changes. Although PyFITS will continue to be
released as a separate package in the near term, the long term plan is to
discontinue PyFITS releases in favor of Astropy. It is expected that direct
support of PyFITS will end mid-2014, so users of PyFITS should plan to make
suitable changes to support the eventual transition to Astropy.

Becoming integrated with Astropy as the \texttt{astropy.io.fits} sub-package
will greatly enhance future development on the existing PyFITS code base in
several areas. First and perhaps foremost is integration with Astropy's
\texttt{Table} interface (\S\ref{sec:table}) which is much more flexible and
powerful than PyFITS' current table interface. We will also be able to
integrate Astropy's unit support (\S\ref{sec:units_main}) in order to attach
units to FITS table columns as well as header values that specify units in
their comments in accordance with the FITS standard. Finally, as the PyWCS
package has also been integrated into Astropy as \texttt{astropy.wcs}
(\S\ref{sec:wcs}) tighter association between data from FITS files and their
world coordinate system (WCS) will be possible.

\subsubsection{ASCII table formats}

\label{sec:ascii}

The \texttt{astropy.io.ascii} sub-package (formerly the standalone project
\texttt{asciitable}\footnote{\url{https://asciitable.readthedocs.org}})
provides the ability to read and write tabular data for a wide variety of
ASCII-based formats. In addition to generic formats such as space-delimited,
tab-delimited or comma-separated values, \texttt{astropy.io.ascii} provides
classes for specialized table formats such as
CDS,\footnote{\url{http://vizier.u-strasbg.fr/doc/catstd.htx}}
IPAC,\footnote{\url{http://irsa.ipac.caltech.edu/applications/DDGEN/}}
IRAF DAOphot \citep{daophot}, and LaTeX. Also included is a flexible class for handling a wide
variety of fixed-width table formats. Finally, this sub-package is designed to be
extensible, making it easy for users to define their own readers and writers for
any other ASCII formats.

\subsubsection{Virtual Observatory tables}

\label{sec:votable}

The \texttt{astropy.io.votable} sub-package (formerly the standalone
project \texttt{vo.table}) provides full support for reading and
writing VOTable format files versions 1.1, 1.2, and the proposed 1.3
\citep{ochsenbein2004votable,ochsenbein2009votable}. It efficiently
stores the tables in memory as \gls{numpy} structured arrays. The file
is read using streaming to avoid reading in the entire file at once
and greatly reducing the memory footprint. VOTable files compressed
using the gzip and bzip2 algorithms are supported transparently, as are
VOTable files where the table data is stored in an external FITS file.

It is possible to convert any one of the tables in a VOTable file to a
\texttt{Table} object (\S\ref{sec:table}), where it can be edited and then
written back to a VOTable file without any loss of data.

The VOTable standard is not strictly adhered to by all VOTable file writers in
the wild. Therefore, \texttt{astropy.io.votable} provides a number of tricks
and workarounds to support as many VOTable sources as possible, whenever the
result would not be ambiguous. A validation tool (\texttt{volint}) is also
provided that outputs recommendations to improve the standard compliance of a
given file, as well as validate it against the official VOTable schema.

\subsection{World Coordinate Systems}

\label{sec:wcs}

The \texttt{astropy.wcs} sub-package contains utilities for managing World
Coordinate System (WCS) transformations in FITS files. These transformations
map the pixel locations in an image to their real-world units, such as their
position on the celestial sphere. This library is specific to WCS as it relates
to FITS as described in the FITS WCS papers
\citep{greisen2002wcs,calabretta2002wcs,greisen2006wcs} and is distinct from a
planned Astropy package that will handle WCS transformations in general,
regardless of their representation.

This sub-package is a wrapper around the \texttt{wcslib} library.\footnote{\url{http://www.atnf.csiro.au/people/mcalabre/WCS/}} Since all of the FITS header parsing is done
using \texttt{wcslib}, it is assured the same behavior as the many other tools
that use \texttt{wcslib}. On top of the basic FITS WCS support, it adds support
for the Simple Imaging Polynomial (SIP) convention and table lookup distortions
\citep{calabretta_sip,sip}. Each of
these transformations can be used independently or together in a fixed
pipeline. The \texttt{astropy.wcs} sub-package also serves as a useful FITS WCS validation
tool, as it is able to report on many common mistakes or deviations from the
standard in a given FITS file.

As mentioned above, the long-term plan is to build a ``generalized'' WCS for mapping
world coordinates to image coordinates (and vice versa).  While only in early
planning stages, such a package would aim to not be tied to the FITS representation
used for the current \texttt{astropy.wcs}.  Such a package would also include
closer connection to other parts of Astropy, for example \texttt{astropy.coordinates} (\S\ref{sec:coordinates}).

\subsection{Cosmology}

\label{sec:cosmology}

The \texttt{astropy.cosmology} sub-package contains classes for representing
widely used cosmologies, and functions for calculating quantities that depend
on a cosmological model. It also contains a framework for working with less
frequently employed cosmologies that may not be flat, or have a time-varying
pressure to density ratio, $w$, for dark energy. The quantities that can be
calculated are generally taken from those described by \citet{Hogg99}. Some
examples are the angular diameter distance, comoving distance, critical
density, distance modulus, lookback time, luminosity distance, and Hubble
parameter as a function of redshift.

\begin{figure}
\center
\caption{Cosmology utilities.\label{code:cosmology}}
\vspace{0.1in}
\fbox{\includegraphics[scale=0.82,clip]{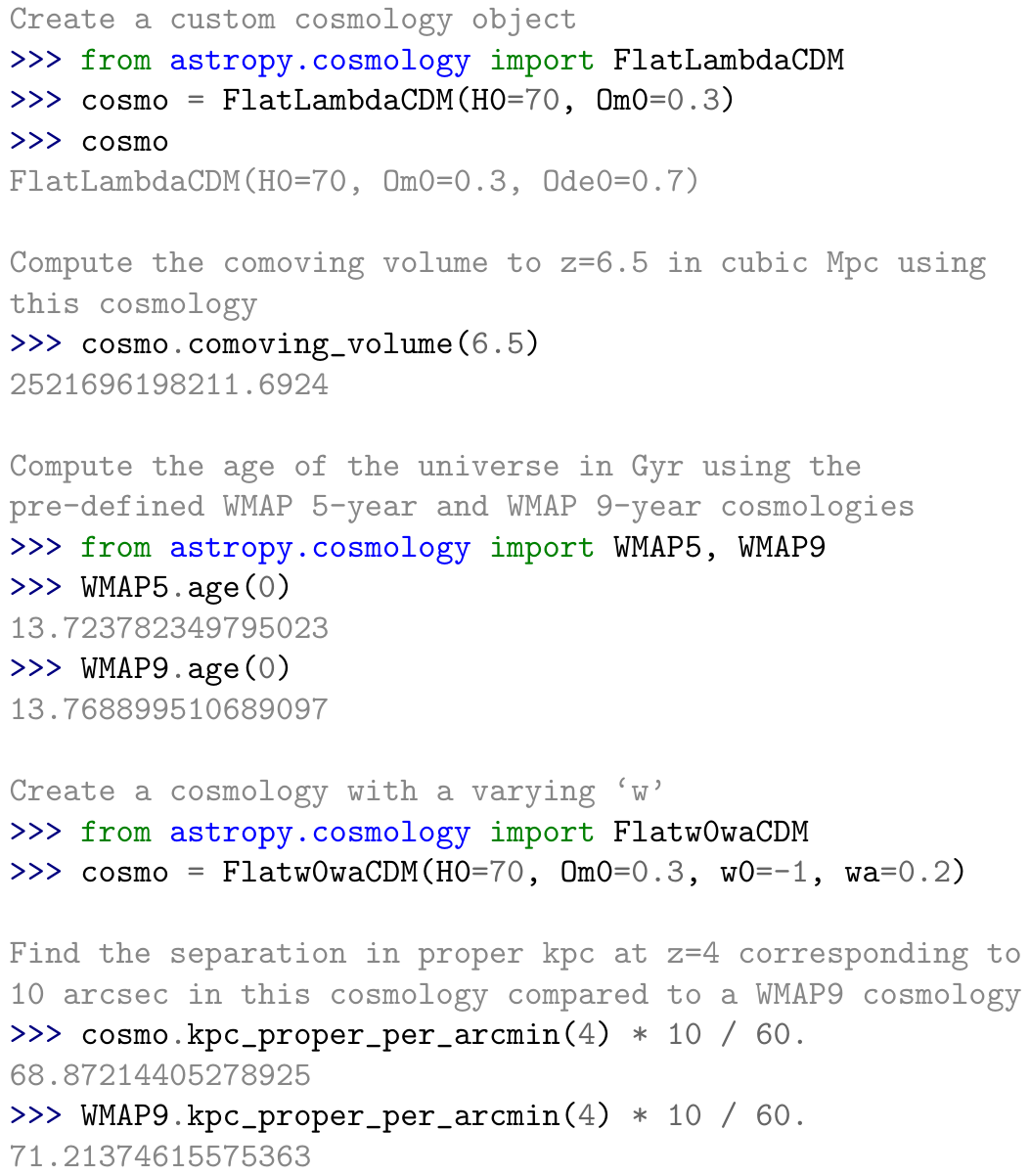}}
\end{figure}

The fundamental model for this sub-package is that any given cosmology is
represented by a class. An instance of this class has attributes giving all the
parameters required to specify the cosmology uniquely, such as the Hubble
parameter, CMB temperature and the baryonic, cold dark matter, and dark energy
densities at $z=0$. One can then use methods of this class to perform
calculations using these parameters.

Figure~\ref{code:cosmology} shows how the \texttt{FlatLambdaCDM} class can be used to
create an object representing a flat $\Lambda$CDM cosmology, and how the
methods of this object can be called to calculate the comoving volume, age and
transverse separation at a given redshift. Further calculations can be performed using the
many methods of the cosmology object as described in the Astropy documentation.
For users who are more comfortable using a procedural coding style, these
methods are also available as functions that take a cosmology class instance as
a keyword argument.

The sub-package provides several pre-defined cosmology instances corresponding
to commonly used cosmological parameter sets. Currently parameters from the
WMAP 5-year \citep{Komatsu09}, 7-year \citep{Komatsu11} and 9-year results
\citep{Hinshaw13} are included (the \texttt{WMAP5}, \texttt{WMAP7}, and
\texttt{WMAP9} classes). The parameters from the \textit{Planck} results
\citep{Planck13} will be included in the next release of Astropy. There are
several classes corresponding to non-flat
cosmologies, and the most common dark energy models are supported: a
cosmological constant, constant $w$, and $w(a) = w_0 + w_a (1-a)$ (e.g.
\citealt{Linder03}, here $a$ is the scale factor). Figure~\ref{code:cosmology}
gives examples showing how to use the pre-defined cosmologies, and how to
define a new cosmology with a time-varying dark energy $w(a)$. Any other
arbitrary cosmology can be represented by sub-classing one of the basic
cosmology classes.

All of the code in the sub-package is tested against the web-based cosmology
calculator by \citet{Wright06} and two other widely-used
calculators.\footnote{\url{http://www.kempner.net/cosmic.php}}$^,$\footnote{\url
{http://www.icosmos.co.uk}} In cases when these calculators are not
precise enough to enable a meaningful comparison, the code is tested against
calculations performed with \textsc{Mathematica}.

\section{Development Approach}

\label{sec:workflow}

A primary guiding philosophy of Astropy is that it is developed for and (at least in part)
\emph{by} the astronomy user community.  This ensures the interface is
designed with the workflow of working astronomers in mind.  At the same time, it aims
to make use of the expertise of software developers to design code that encourages
good software practices such as a consistent and clean API, thorough documentation,
and integrated testing.  It is also dedicated to remaining open source to enable wide adoption
and render input from all users easier, and is thus
released with a 3-clause BSD-style license (A license of this sort is ``permissive''
in that it allows usage of the code for any purposes as long as notice of the Astropy copyright
and disclaimers of warranty are given). Achieving these aims requires code collaboration
between over 30 geographically-distributed developers, and here we describe
our development workflow with the hope that it may be replicated by other astronomy
software projects that are likely to have similar needs.

To enable this collaboration, we have made use of the
GitHub\footnote{\url{http://www.github.com}} open source code hosting and
development  platform. The main repository for \astropy is stored in a
git\footnote{\url{http://git-scm.com}} repository on GitHub, and any
non-trivial changes are made via \textit{pull requests}, which are a mechanism
for submitting code for review by other developers prior to merging into the
main code base. This workflow aids in increasing the quality, documentation and
testing of the code to be included in \astropy. Not all contributions are
necessarily accepted - community consensus is needed for incorporating major
new functionality in \astropy, and any new feature has to be justified to avoid
implementing features that are only useful to a minority of users, but may
cause issues in the future.

At the time of writing, \astropy includes several thousand tests, which are
small units of code that check that functions, methods, and classes in \astropy
are behaving as expected, both in terms of scientific correctness and from a
programming interface perspective. We make use of \textit{continuous
integration}, which is the process of running all the tests under various
configurations (such as different versions of Python or \gls{numpy}, and on different
platforms) in order to ensure that the package is held to the highest standard
of stability. In particular, any change made via a pull request is subject to
extensive testing before being merged into the core repository. For the latter,
we make use of Travis,\footnote{\url{https://travis-ci.org}} while for running
more extensive tests across Linux, MacOS X, and Windows, we make use of
Jenkins\footnote{\url{http://jenkins-ci.org}} (both are examples of continuous
integration systems).

This development workflow has worked very well so far, allowing contributions
by many developers, and blurring the line between developers and users. Indeed,
users who encounter bugs and who know how to fix them can submit suggested
changes. We have also implemented a feature that means that anyone reading the
documentation at \url{http://docs.astropy.org} can suggest improvements to the
documentation with just a few clicks in a web browser without any prior
knowledge of the git version control system.

\section{Planned functionality}

\label{sec:future}

Development on the Astropy package is very active, and in addition to
some of the incremental improvements to existing sub-packages described in the
text, we are focusing on implementing major new functionality in
several areas for the next (v0.3) release (some of which have already been
implemented in the publicly-available developer version):

\begin{itemize}
\item Improving interoperability between packages, which includes for example seamlessly integrating the \texttt{astropy.units} framework across all sub-packages
\item Adding support for NumPy arrays in the coordinates sub-package, which will allow the efficient representation and conversions of coordinates in large datasets
\item Supporting more file formats for reading and writing \texttt{Table} and \texttt{NDData} objects
\item Implementing a Virtual Observatory cone search tool \citep{vo-cone-search}
\item Implementing a generalized model-fitting framework
\item Implementing statistical functions commonly used in Astronomy
\end{itemize}

In the longer term, we are already planning the following major functionality:

\begin{itemize}
\item Image analysis tools, including aperture and point spread function (PSF) photometry
\item Spectroscopic analysis tools
\item Generalized WCS transformations beyond the FITS WCS standard
\item A SAMP server/client (ported from the SAMPy\footnote{\url{http://pythonhosted.org/sampy}} package)
\item Support for the Simple Image Access Protocol (SIAP) \citep{vo-siap}
\item Support for the Table Access Protocol (TAP) \citep{vo-tap} is under consideration
\end{itemize}

\noindent and undoubtedly the core functionality will grow beyond this. In
fact, the \astropy package will likely remain a continuously-evolving package,
and will thus never be considered `complete' in the traditional sense.

\section{Summary}

\label{sec:summary}

We have presented the first public release of the Astropy package (v0.2), a
core Python package for astronomers. In this paper we have described the main
functionality in this release, which includes:

\begin{itemize}
\item Units and unit conversions (\S\ref{sec:units_main})
\item Absolute dates and times (\S\ref{sec:time})
\item Celestial coordinate systems (\S\ref{sec:coordinates})
\item Tabular and gridded data (\S\ref{sec:table})
\item Support for common astronomical file formats (\S\ref{sec:io})
\item World Coordinate System transformations (\S\ref{sec:wcs})
\item Cosmological calculations (\S\ref{sec:cosmology}).
\end{itemize}

We also briefly described our development approach (\S\ref{sec:workflow}),
which has enabled an international collaboration of scientists and software
developers to create and contribute to the package. We outlined our plans for
the future (\S\ref{sec:future}) which includes more interoperability of
sub-packages, as well as new functionality.

We invite members of the community to join the effort by adopting the Astropy package for
their own projects, reporting any issues, and whenever possible, developing new
functionality.

\section*{Acknowledgements}

\label{sec:acknowledgements}

We thank the referee, Igor Chiligarian, for suggestions that helped improve this paper.
We would like to thank the \gls{numpy}, \gls{scipy}, \gls{ipython} and
\gls{matplotlib} communities for providing their packages which are invaluable
to the development of Astropy. We thank the GitHub team for providing us with
an excellent free development platform. We also are grateful to Read the Docs
(\url{https://readthedocs.org/}), Shining Panda
(\url{https://www.shiningpanda-ci.com/}), and Travis
(\url{https://www.travis-ci.org/}) for providing free documentation
hosting and testing respectively. Finally, we would like to thank all the
\astropy users that have provided feedback and submitted bug reports.
The contribution by T. Aldcroft was funded by NASA contract NAS8-39073.
The name resolution functionality shown in
Figure~\ref{fig:code_coordinates} makes use of the SIMBAD database, operated
at CDS, Strasbourg, France.

\bibliographystyle{apj_custom}
\bibliography{apj-jour,references}

\begin{thebibliography}{40}
\expandafter\ifx\csname natexlab\endcsname\relax\def\natexlab#1{#1}\fi

\bibitem[{{Barrett} \& {Bridgman}(1999)}]{barrett1999pyfits}
{Barrett}, P.~E., \& {Bridgman}, W.~T. 1999, in Astronomical Society of the
  Pacific Conference Series, Vol. 172, Astronomical Data Analysis Software and
  Systems VIII, 483

\bibitem[{Behnel {et~al.}(2011)Behnel, Bradshaw, Citro, Dalcin, Seljebotn, \&
  Smith}]{cython}
Behnel, S., Bradshaw, R., Citro, C., Dalcin, L., Seljebotn, D., \& Smith, K.
  2011, Computing in Science Engineering, 13, 31

\bibitem[{{Berry} \& {Jenness}(2012)}]{pyast}
{Berry}, D.~S., \& {Jenness}, T. 2012, in Astronomical Society of the Pacific
  Conference Series, Vol. 461, Astronomical Data Analysis Software and Systems
  XXI, ed. P.~{Ballester}, D.~{Egret}, \& N.~P.~F. {Lorente}, 825

\bibitem[{Calabretta \& Greisen(2002)}]{calabretta2002wcs}
Calabretta, M.~R., \& Greisen, E.~W. 2002, Astronomy \& Astrophysics, 1077

\bibitem[{{Calabretta} {et~al.}(2004){Calabretta}, {Valdes}, {Greisen}, \&
  {Allen}}]{calabretta_sip}
{Calabretta}, M.~R., {Valdes}, F., {Greisen}, E.~W., \& {Allen}, S.~L. 2004, in
  Astronomical Society of the Pacific Conference Series, Vol. 314, Astronomical
  Data Analysis Software and Systems (ADASS) XIII, ed. F.~{Ochsenbein}, M.~G.
  {Allen}, \& D.~{Egret}, 551

\bibitem[{Derriere {et~al.}(2012)Derriere, Gray, Louys, McDowell, Ochsenbein,
  Osuna, Rino, \& Salgado}]{derriere2012vounit}
Derriere, S., Gray, N., Louys, M., McDowell, J., Ochsenbein, F., Osuna, P.,
  Rino, B., \& Salgado, J. 2012, Units in the VO, Version 1.0, {IVOA Proposed
  Recommendation 20 August 2012} edn.

\bibitem[{George \& Angelini(1995)}]{george1995ogip}
George, I., \& Angelini, L. 1995, Specification of Physical Units within OGIP
  (Office of Guest Investigator Programs) FITS files

\bibitem[{{Greenfield}(2011)}]{greenfield11}
{Greenfield}, P. 2011, in Astronomical Society of the Pacific Conference
  Series, Vol. 442, Astronomical Data Analysis Software and Systems XX, ed.
  I.~N. {Evans}, A.~{Accomazzi}, D.~J. {Mink}, \& A.~H. {Rots}, 425

\bibitem[{Greisen \& Calabretta(2002)}]{greisen2002wcs}
Greisen, E.~W., \& Calabretta, M.~R. 2002, Astronomy \& Astrophysics, 1061

\bibitem[{Greisen {et~al.}(2006)Greisen, Calabretta, Valdes, \&
  Allen}]{greisen2006wcs}
Greisen, E.~W., Calabretta, M.~R., Valdes, F.~G., \& Allen, S.~L. 2006,
  Astronomy \& Astrophysics, 747

\bibitem[{{Guinot} \& {Seidelmann}(1988)}]{guinot88}
{Guinot}, B., \& {Seidelmann}, P.~K. 1988, \aap, 194, 304

\bibitem[{{Hinshaw} {et~al.}(2012){Hinshaw}, {Larson}, {Komatsu}, {Spergel},
  {Bennett}, {Dunkley}, {Nolta}, {Halpern}, {Hill}, {Odegard}, {Page}, {Smith},
  {Weiland}, {Gold}, {Jarosik}, {Kogut}, {Limon}, {Meyer}, {Tucker}, {Wollack},
  \& {Wright}}]{Hinshaw13}
{Hinshaw}, G. {et~al.} 2012, arXiv:1212.5226

\bibitem[{{Hogg}(1999)}]{Hogg99}
{Hogg}, D.~W. 1999, ArXiv Astrophysics e-prints, arXiv:astro-ph/9905116

\bibitem[{Jenness \& Berry(2013)}]{palpy}
Jenness, T., \& Berry, D.~S. 2013, in ASP Conf Ser., Vol. TBD, ADASS XXII, ed.
  D.~Friedel, M.~Freemon, \& R.~Plante (San Francisco: ASP), in press

\bibitem[{Jones {et~al.}(2001)Jones, Oliphant, \& Peterson}]{jones2001scipy}
Jones, E., Oliphant, T., \& Peterson, P. 2001, http://www. scipy. org/

\bibitem[{{Kaplan}(2005)}]{usnocircular179}
{Kaplan}, G.~H. 2005, U.S.~Naval Observatory Circulars, 179

\bibitem[{{Komatsu} {et~al.}(2009){Komatsu}, {Dunkley}, {Nolta}, {Bennett},
  {Gold}, {Hinshaw}, {Jarosik}, {Larson}, {Limon}, {Page}, {Spergel},
  {Halpern}, {Hill}, {Kogut}, {Meyer}, {Tucker}, {Weiland}, {Wollack}, \&
  {Wright}}]{Komatsu09}
{Komatsu}, E. {et~al.} 2009, \apjs, 180, 330

\bibitem[{{Komatsu} {et~al.}(2011){Komatsu}, {Smith}, {Dunkley}, {Bennett},
  {Gold}, {Hinshaw}, {Jarosik}, {Larson}, {Nolta}, {Page}, {Spergel},
  {Halpern}, {Hill}, {Kogut}, {Limon}, {Meyer}, {Odegard}, {Tucker}, {Weiland},
  {Wollack}, \& {Wright}}]{Komatsu11}
------. 2011, \apjs, 192, 18

\bibitem[{{Kovalevsky}(2001)}]{kovalevsky01}
{Kovalevsky}, J. 2001, in Journ{\'e}es 2000 - syst{\`e}mes de r{\'e}f{\'e}rence
  spatio-temporels. J2000, a fundamental epoch for origins of reference systems
  and astronomical models, ed. N.~{Capitaine}, 218--224

\bibitem[{{Linder}(2003)}]{Linder03}
{Linder}, E.~V. 2003, Physical Review Letters, 90, 091301

\bibitem[{{Louys} {et~al.}(2011){Louys}, {Bonnarel}, {Schade}, {Dowler},
  {Micol}, {Durand}, {Tody}, {Michel}, {Salgado}, {Chilingarian}, {Rino},
  {Santander-Vela}, \& {Skoda}}]{vo-tap}
{Louys}, M. {et~al.} 2011, arXiv:1111.1758

\bibitem[{McKinney(2012)}]{pandas}
McKinney, W. 2012, Python for Data Analysis (O'Reilly Media, Incorporated)

\bibitem[{Ochsenbein(2000)}]{ochsenbein2000cds}
Ochsenbein, F. 2000, Astronomical Catalogues and Tables Adopted Standards,
  Version 2.0

\bibitem[{Oliphant(2006)}]{oliphant2006guide}
Oliphant, T. 2006, A Guide to NumPy, Vol.~1 (Trelgol Publishing USA)

\bibitem[{Oschenbein {et~al.}(2004)Oschenbein, Williams, Davenhall, Durand,
  Fernique, Giaretta, Hanisch, McGlynn, Szalay, Taylor, \&
  Wicenec}]{ochsenbein2004votable}
Oschenbein, F. {et~al.} 2004, VOTable Format Definition, Version 1.1,
  International Virtual Observatory Alliance (IVOA)

\bibitem[{Oschenbein {et~al.}(2009)Oschenbein, Williams, Davenhall, Durand,
  Fernique, Giaretta, Hanisch, McGlynn, Szalay, Taylor, \&
  Wicenec}]{ochsenbein2009votable}
------. 2009, VOTable Format Definition, Version 1.2, International Virtual
  Observatory Alliance (IVOA)

\bibitem[{{Pence} {et~al.}(2010){Pence}, {Chiappetti}, {Page}, {Shaw}, \&
  {Stobie}}]{fits3}
{Pence}, W.~D., {Chiappetti}, L., {Page}, C.~G., {Shaw}, R.~A., \& {Stobie}, E.
  2010, \aap, 524, A42

\bibitem[{{Planck Collaboration} {et~al.}(2013){Planck Collaboration}, {Ade},
  {Aghanim}, {Armitage-Caplan}, {Arnaud}, {Ashdown}, {Atrio-Barandela},
  {Aumont}, {Baccigalupi}, {Banday}, \& et~al.}]{Planck13}
{Planck Collaboration} {et~al.} 2013, arXiv:1303.5076

\bibitem[{{Pontzen} {et~al.}(2013){Pontzen}, {Ro{\v s}kar}, {Stinson}, {Woods},
  {Reed}, {Coles}, \& {Quinn}}]{pynbody}
{Pontzen}, A., {Ro{\v s}kar}, R., {Stinson}, G.~S., {Woods}, R., {Reed}, D.~M.,
  {Coles}, J., \& {Quinn}, T.~R. 2013, {pynbody: Astrophysics Simulation
  Analysis for Python}, Astrophysics Source Code Library, ascl:1305.002

\bibitem[{{Rhodes}(2011)}]{pyephem}
{Rhodes}, B.~C. 2011, {PyEphem: Astronomical Ephemeris for Python},
  Astrophysics Source Code Library, ascl:1112.014

\bibitem[{{Shupe} {et~al.}(2005){Shupe}, {Moshir}, {Li}, {Makovoz}, {Narron},
  \& {Hook}}]{sip}
{Shupe}, D.~L., {Moshir}, M., {Li}, J., {Makovoz}, D., {Narron}, R., \& {Hook},
  R.~N. 2005, in Astronomical Society of the Pacific Conference Series, Vol.
  347, Astronomical Data Analysis Software and Systems XIV, ed. P.~{Shopbell},
  M.~{Britton}, \& R.~{Ebert}, 491

\bibitem[{{Soffel} {et~al.}(2003){Soffel}, {Klioner}, {Petit}, {Wolf},
  {Kopeikin}, {Bretagnon}, {Brumberg}, {Capitaine}, {Damour}, {Fukushima},
  {Guinot}, {Huang}, {Lindegren}, {Ma}, {Nordtvedt}, {Ries}, {Seidelmann},
  {Vokrouhlick{\'y}}, {Will}, \& {Xu}}]{soffel03}
{Soffel}, M. {et~al.} 2003, \aj, 126, 2687

\bibitem[{{Stetson}(1987)}]{daophot}
{Stetson}, P.~B. 1987, \pasp, 99, 191

\bibitem[{{Terlouw} \& {Vogelaar}(2012)}]{kapteyn}
{Terlouw}, J.~P., \& {Vogelaar}, M.~G.~R. 2012, {Kapteyn Package, version 2.2},
  {Kapteyn Astronomical Institute}, Groningen

\bibitem[{{Tody} {et~al.}(2011){Tody}, {Plante}, \& {Harrison}}]{vo-siap}
{Tody}, D., {Plante}, R., \& {Harrison}, P. 2011, arXiv:1110.0499

\bibitem[{{Tollerud}(2012)}]{astropysics}
{Tollerud}, E. 2012, {Astropysics: Astrophysics utilities for python},
  Astrophysics Source Code Library, ascl:1207.007

\bibitem[{Van Der~Walt {et~al.}(2011)Van Der~Walt, Colbert, \&
  Varoquaux}]{van2011numpy}
Van Der~Walt, S., Colbert, S., \& Varoquaux, G. 2011, Computing in Science \&
  Engineering, 13, 22

\bibitem[{{Wallace}(2011)}]{sofa_wallace}
{Wallace}, P.~T. 2011, Metrologia, 48, 200

\bibitem[{{Williams} {et~al.}(2011){Williams}, {Hanisch}, {Szalay}, \&
  {Plante}}]{vo-cone-search}
{Williams}, R., {Hanisch}, R., {Szalay}, A., \& {Plante}, R. 2011,
  arXiv:1110.0498

\bibitem[{{Wright}(2006)}]{Wright06}
{Wright}, E.~L. 2006, \pasp, 118, 1711

\end{thebibliography}

\end{document}